\documentclass[%
twocolumn,
showpacs,
nofootinbib,
nobibnotes,
amsmath,amssymb,
aps,
pra,
floatfix,
]{revtex4}
\usepackage{graphicx}
\usepackage{dcolumn}
\usepackage{bm}
\usepackage{hyperref}
\usepackage{natbib}
\usepackage{subfigure}
\usepackage{color}

\begin{document}

\preprint{APS/123-QED}

\title{Dynamics of mode entanglement in a system of cavities\\ coupled with a chiral mirror}
\author{Ali \"{U}. C. Hardal}
\email{ahardal@ku.edu.tr}
\affiliation{Department of Physics, Ko\c{c} University, \.Istanbul, 34450, Turkey}
\affiliation{Department of Electrical Engineering, Princeton University, Princeton, New Jersey 08544, USA}
%
%
\begin{abstract}
We investigate the Hermitian and the non-Hermitian dynamics of the mode entanglement in two identical optical cavities coupled by a chiral mirror. By employing the non-Hermitian quantum evolution, we calculate the logarithmic negativity measure of entanglement for initially Fock, coherent and squeezed states, separately. We verify the non-conservation of mean spin for the initially coherent and squeezed states when the coupling is non-reciprocal and report the associated spin noise for each case. We examine the effects of non-conserved symmetries on the mode correlations and determine the degree of non-reciprocal coupling to establish robust quantum entanglement.
\end{abstract}
\maketitle
\section{Introduction}
The interest in the systems which exhibits non-Hermitian quantum mechanical interactions~\cite{bender1998real,bender1999pt,mostafazadeh2002pseudo,mostafazadeh2003exact,bender2007making} has been intensified particularly in the last decade. They have been reported in many research fields including soliton-plasmon systems~\cite{karakaya,milian2012soliton,ferrando2013variational}, hybridized metamaterials~\cite{PhysRevA.87.053824}, coupled microcavities~\cite{peng2014parity}, waveguides~\cite{PhysRevLett.101.080402,PhysRevA.87.013816}, optical lattices~\cite{PhysRevLett.110.223902,longhi2014pt} and Bose-Einstein condensates~\cite{graefe2008non,graefe2008mean,dast2013eigenvalue}. $\mathcal{PT}$-symmetric lasers and anti-lasers~\cite{PhysRevA.82.031801,PhysRevLett.105.053901,PhysRevLett.108.173901,PhysRevA.84.063833}, cloaking devices~\cite{PhysRevA.82.043803} and unidirectional invisibility~\cite{mostafazadeh2013invisibility,PhysRevLett.106.213901} constitute some of the intriguing implementations of such quantum mechanical systems.

Along a similar direction, a generic quantum optical model of coupled resonators which exhibits non-Hermiticity has been proposed very recently~\cite{santos2012non}. In the model, two independent quantum oscillators are weakly coupled with a chiral mirror. The dynamical analysis by the usual Heisenberg approach has revealed the asymmetric photon exchange between the resonators. In addition, the non-conservation of the total photon number was reported for the same model by employing the non-Hermitian quantum evolution~\cite{karakaya}. The non-conservation of the mean number of photons is an interplay between the quantum coherence and the non-Hermitian dynamics. Here, we aim to reveal whether there is relation between the mean spin and the entanglement dynamics as well. 

In the present contribution, we consider the model system proposed in Ref.~\cite{santos2012non}. We investigate the non-Hermitian dynamics of the mode entanglement by the means of the logarithmic negativity measure~\cite{plenio2005logarithmic}. Mode entanglement occurs in the second quantization picture~\cite{PhysRevA.72.064306,cunha2007entanglement} and can be witnessed via the covariances of the two distinct modes~\cite{PhysRevLett.96.050503,PhysRevA.74.032333}. Furthermore, the relations between mode correlations and the spin noise in coupled cavity systems have been revealed only recently~\cite{hardal2013spin}. The absence of bipartite mode entanglement due to the lack of nonlinearity in the model system under consideration has also been verified~\cite{karakaya}. Here, we report the existence of genuine mode entanglement in the generic model that are robust and controllable via the asymmetry in the coupling of the two modes.

In our numerical analysis, we assume that one cavity is in its vacuum while the other in a Fock, coherent and squeezed state, separately. The non-conservation of the mean spin is verified for the initially coherent and squeezed states when the coupling between the cavities is non-reciprocal. The associated spin noise measured by the variances of the corresponding spin operators of the coupled resonators is reported. We find that the mode entanglement is more robust if the system is in a coherent state and non-Hermitian, though it displays high amplitude oscillations in comply with the noise dynamics. We, then, consider an initially single-mode squeezed state to compensate spin noise and clarify the interference between population, spin and entanglement dynamics.

In a recent contribution~\cite{hardal2013spin}, we investigated a more general set up consists of two nonlinear cavities coupled either with single- or two-photon exchange interactions. Quantum entanglement and field coherence were investigated in the steady state in a comparative manner. The focus of the work was to reveal profound relations between coherence, localization (delocalization) of photons and quantum correlations. Here, we consider a more fundamental model which exhibits non-Hermitian dynamics. Our motivation is to dynamically investigate the modal entanglement and its response to the broken symmetries due to the asymmetric coupling between the cavities.
   
This paper is organized as follows. In Sec.~\ref{sec:model}, we briefly review the model system and the governing non-Hermitian quantum dynamics. In Sec.~\ref{sec:results}, we present our results and we finally conclude in Sec.~\ref{sec:conc}.
\section{The Model System and The Non-Hermitian Dynamics}\label{sec:model}
We consider two identical  optical cavities, $A$ and $B$, which are coupled by a chiral mirror. The dynamics of the system is governed by the Hamiltonian~\cite{santos2012non}
\begin{equation}\label{eq:model1}
H = \omega_0(a^\dagger a+b^\dagger b)+g_{AB}ab^\dagger+g_{BA}a^\dagger b.
\end{equation}
Here, $a$ and $b$ are the annihilation operators of the cavity modes, $\omega_0$ is the resonant transition frequency for each cavity and  $g_{AB}$ and $g_{BA}$ denote the coupling strengths. 

The Hamiltonian~(\ref{eq:model1}) can equivalently be written as
\begin{equation}\label{eq:model2}
H = \omega_0N+g_{AB}S_{+}+g_{BA}S_{-},
\end{equation}
where we made use of the pseudo-spin operators for the two-resonator system
\begin{eqnarray}\label{eq:spins}
\nonumber S_x &:=&\frac{1}{2}(a^{\dagger}b+b^{\dagger}a),\\
S_y &:=&\frac{-i}{2}(a^{\dagger}b-b^{\dagger}a),\\
\nonumber S_z &:=&\frac{1}{2}(a^{\dagger}a-b^{\dagger}b),
\end{eqnarray}
with $S_{+} :=S_x+iS_y$, $S_{-}:=S_x-iS_y$ and $N=a^{\dagger}a+b^{\dagger}b$. The operators given in Eq.~(\ref{eq:spins}) satisfy the usual spin algebra $[S_{\alpha},S_{\beta}]=\epsilon^{\alpha\beta\gamma}S_{\gamma}$ with $\alpha,\beta,\gamma\in{x,y,z}$ and $\epsilon^{\alpha\beta\gamma}$ is Levi-Civita tensor. In the case that $g_{AB}=g_{BA}$, the model describes a reciprocal, single-photon exchange type coupling between two resonant cavities. The latter type of coupling generally induces genuine mode correlations which can be expressed with the covariances of the two modes~\cite{PhysRevLett.96.050503,PhysRevA.74.032333} and can further be related to the spin noise~\cite{hardal2013spin}. 

When $g_{AB}\neq g_{BA}$ the system becomes non-Hermitian even if the coupling coefficients are real as it can easily be seen from Eq.~(\ref{eq:model2}). The asymmetric coupling between the cavities behaves as a dissipation or an amplification channel depending on which direction that the symmetry is broken, as a result the system does not conserve the mean number of photons $\langle N\rangle$~\cite{karakaya}. However, the non-conservation of the mean number of excitations do result from the applied dynamical approach as well as the initial preparation of the system~\cite{santos2014asymmetrical,santos2012non,karakaya}. Here, we adopt the approach for which the total number of photons $\langle N\rangle$ is not conserved. We also verify that if the initial state of the system is a coherent or a squeezed state, the non-reciprocal dynamics does not conserve the mean spin $\langle S^2\rangle=\langle S_x^2+S_y^2+S_z^2\rangle$ as well.
\begin{figure*}[ht!]
     \begin{center}
            \subfigure[]{%
            \label{fig:fig1a}
            \includegraphics[width=5.85cm]{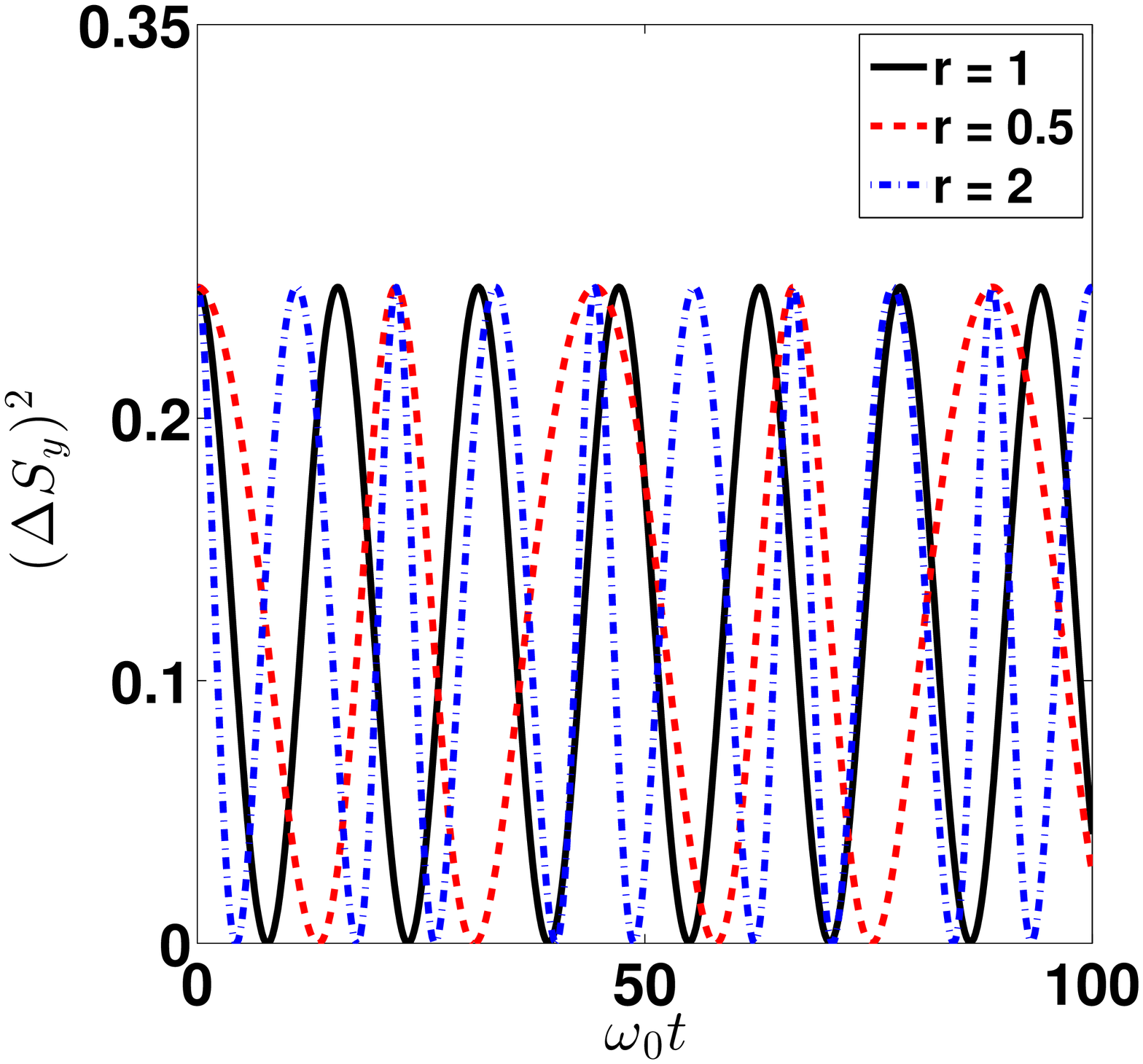}
        }
            \subfigure[]{%
            \label{fig:fig1b}
            \includegraphics[width=5.85cm]{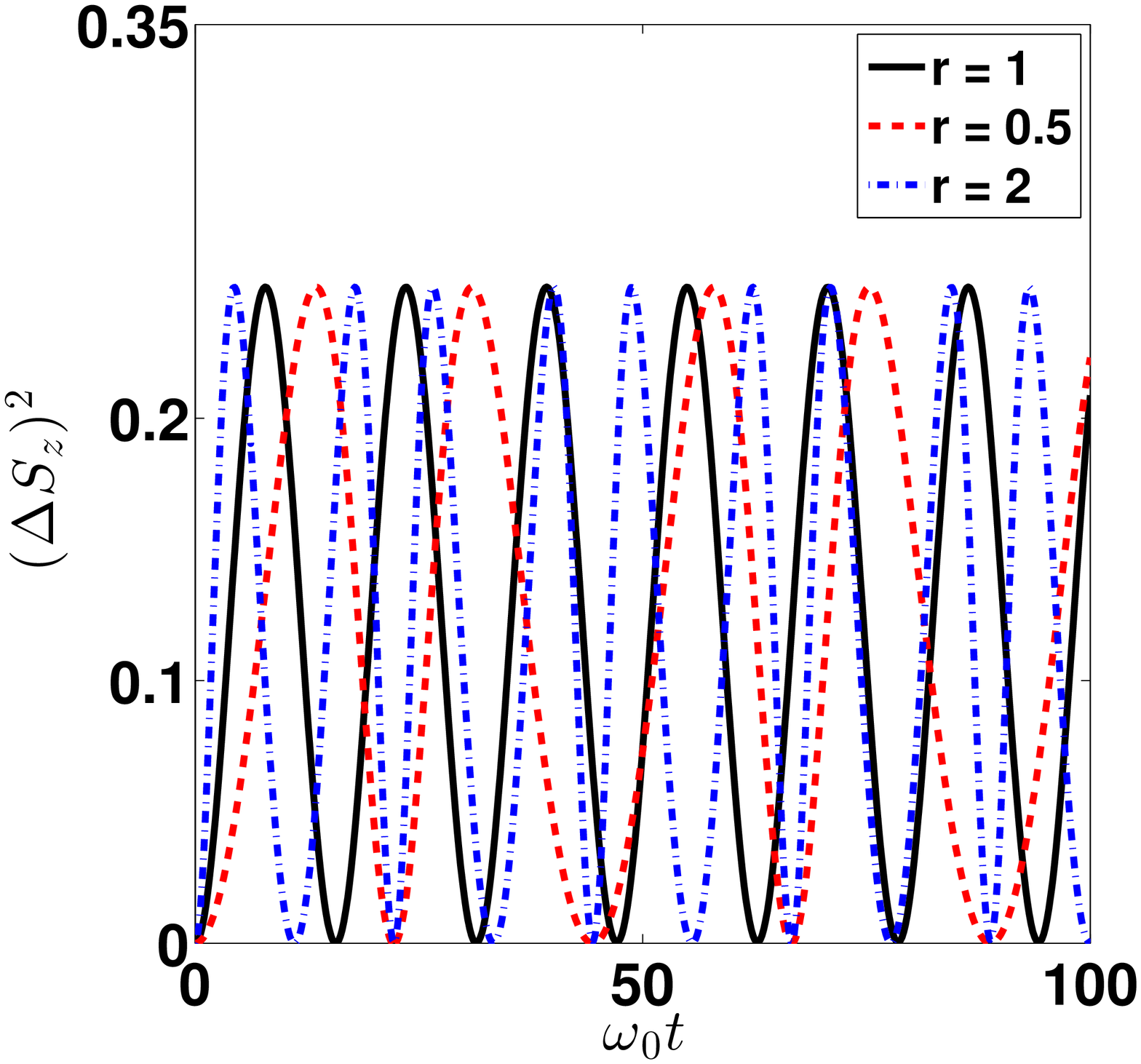}
        }%
	    \subfigure[]{%
            \label{fig:fig1c}
            \includegraphics[width=5.85cm]{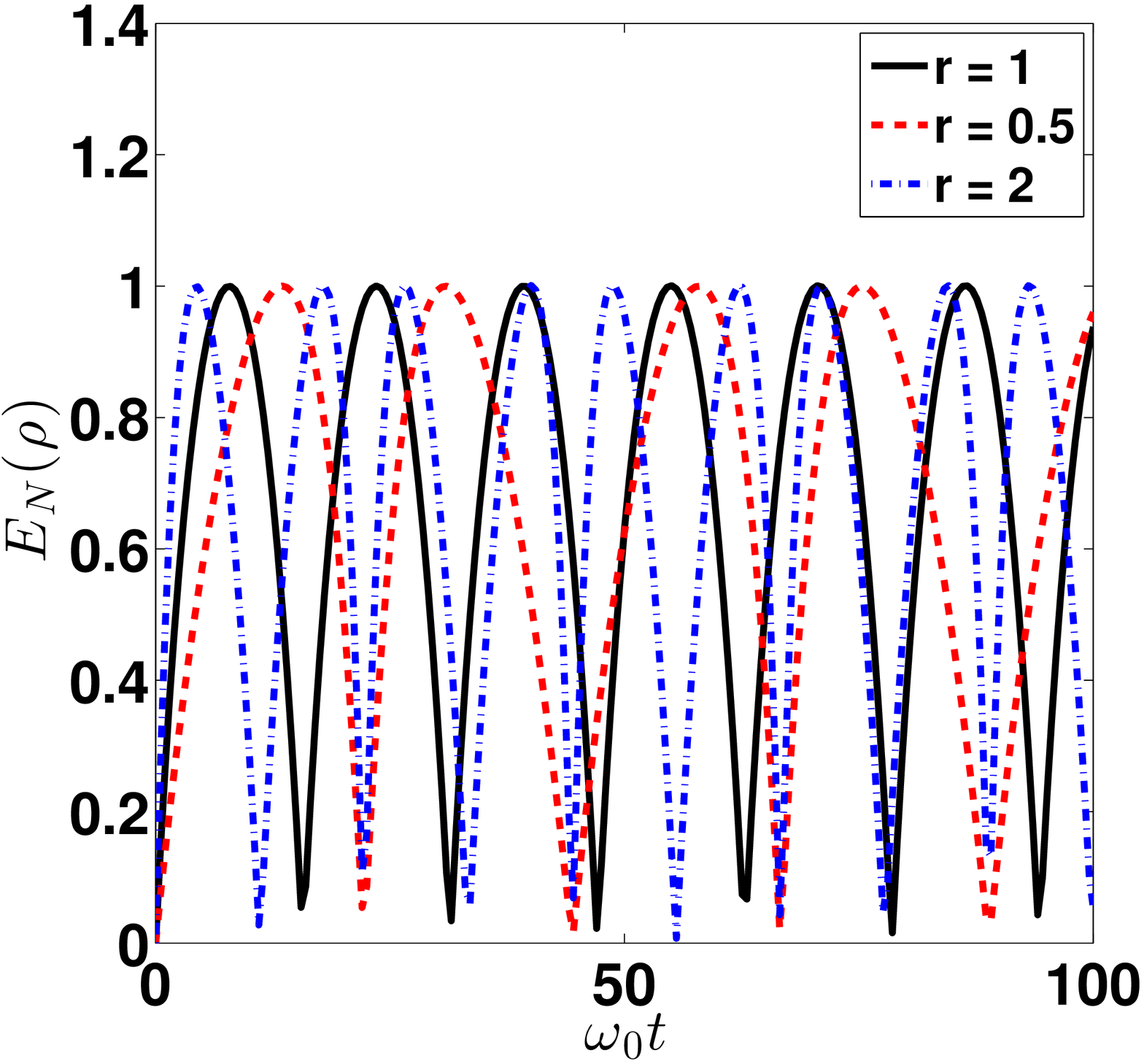}
            }%
    \end{center}
    \caption{%
       Dependence of (a) $(\Delta S_y)^2$, (b) $(\Delta S_z)^2$ and (c) $E_N(\rho)$ for $r=1$ (black-solid), $r = 0.5$ (red-dashed) and $r=2$ (blue-dot-dashed) with respect to the scaled time $\omega_0t$ for an initially Fock state $|\psi(0)\rangle=|1\rangle|0\rangle$.}%
   \label{fig:fig1}
\end{figure*}

The dynamics of the system may be investigated with the usual Heisenberg approach~\cite{santos2012non}, however it has been recently shown~\cite{karakaya} that to capture the effects of non-reciprocal dynamics one should consider a more general formalism~\cite{sergi2013non,graefe2008mean,PhysRevA.42.1467,PhysRevLett.109.230405}. To that end, we first write the Hamiltonian~(\ref{eq:model1}) as the sum of its Hermitian $H_+$ and anti-Hermitian $H_-$ parts
\begin{equation}\label{eq:parts}
H=H_+ + H_-,
\end{equation}
where $H_{\pm}:=1/2(H\pm H^\dagger)$ with $H_{\pm}=\pm H_{\pm}^{\dagger}$. The time evolution of a state $\rho(t)$ of the system can be determined by the modified Liouville-von Neumann master equation~\cite{karakaya}
\begin{equation}\label{eq:master}
\frac{\partial}{\partial t}\rho(t)=-i[H_+,\rho(t)]_+-i[H_-,\rho(t)]_-,
\end{equation}
where $[,]_+$ and $[,]_-$ represent the commutator and the anti-commutator of the corresponding operators. Due to the non-unitary character of the Eq.~(\ref{eq:master}), we renormalize the density operator as
\begin{equation}\label{eq:density}
\rho(t)^\prime:=\frac{\rho(t)}{Tr(\rho(t))}.
\end{equation}
It follows that the expectation value of a given observable $Q$ is calculated via the relation
\begin{equation}\label{eq:expect}
\langle Q\rangle:=\frac{Tr(\rho(t)Q)}{Tr(\rho(t))}.
\end{equation} 
In the following section, we shall first define the measure of quantum entanglement and the parameters that are going to be used in our analysis. We, then, present our results for initially fock, coherent and squeezed states, separately. 
\section{Results and Discussions}\label{sec:results}
Here, we shall discuss the non-Hermitian quantum dynamics of mode entanglement between the two cavity modes. In our numerical analysis, we use the QuTiP: Quantum Toolbox in Python software~\cite{johansson2013qutip}. We set the Hilbert space dimensions of the modes $N_A=N_B=N=25$ which we concluded that is sufficient for the analysis of quantum entanglement. We repeated our calculations up to $N=30$ and obtained the same results. In particular, for the dimensions $N<15$, we found that the results are not stable. We note that the latter bounds on the Hilbert space dimensions are not physical and can be differ with respect to the preferred numerical algorithm and method.

We make our calculations for $g_{AB}=g.r$, $g_{BA}=g$ with $r=0.5,1,2$~\cite{karakaya}. Here, $r=1$ corresponds to the Hermitian whereas $r=0.5,2$ corresponds to non-Hermitian cases, respectively.
\begin{figure*}[ht!]
     \begin{center}
            \subfigure[]{%
            \label{fig:fig2a}
            \includegraphics[width=5.85cm]{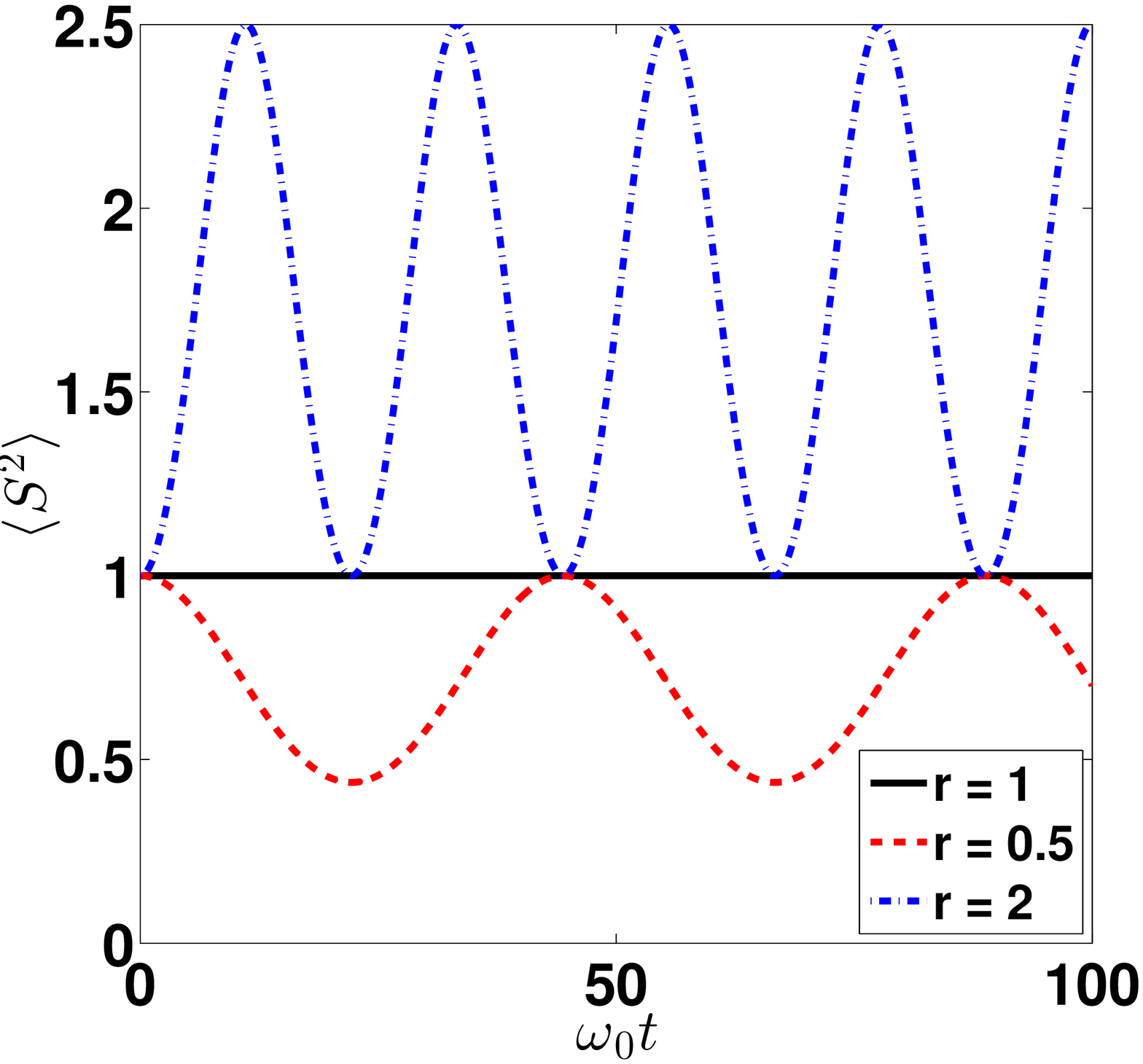}
        }
            \subfigure[]{%
            \label{fig:fig2b}
            \includegraphics[width=5.85cm]{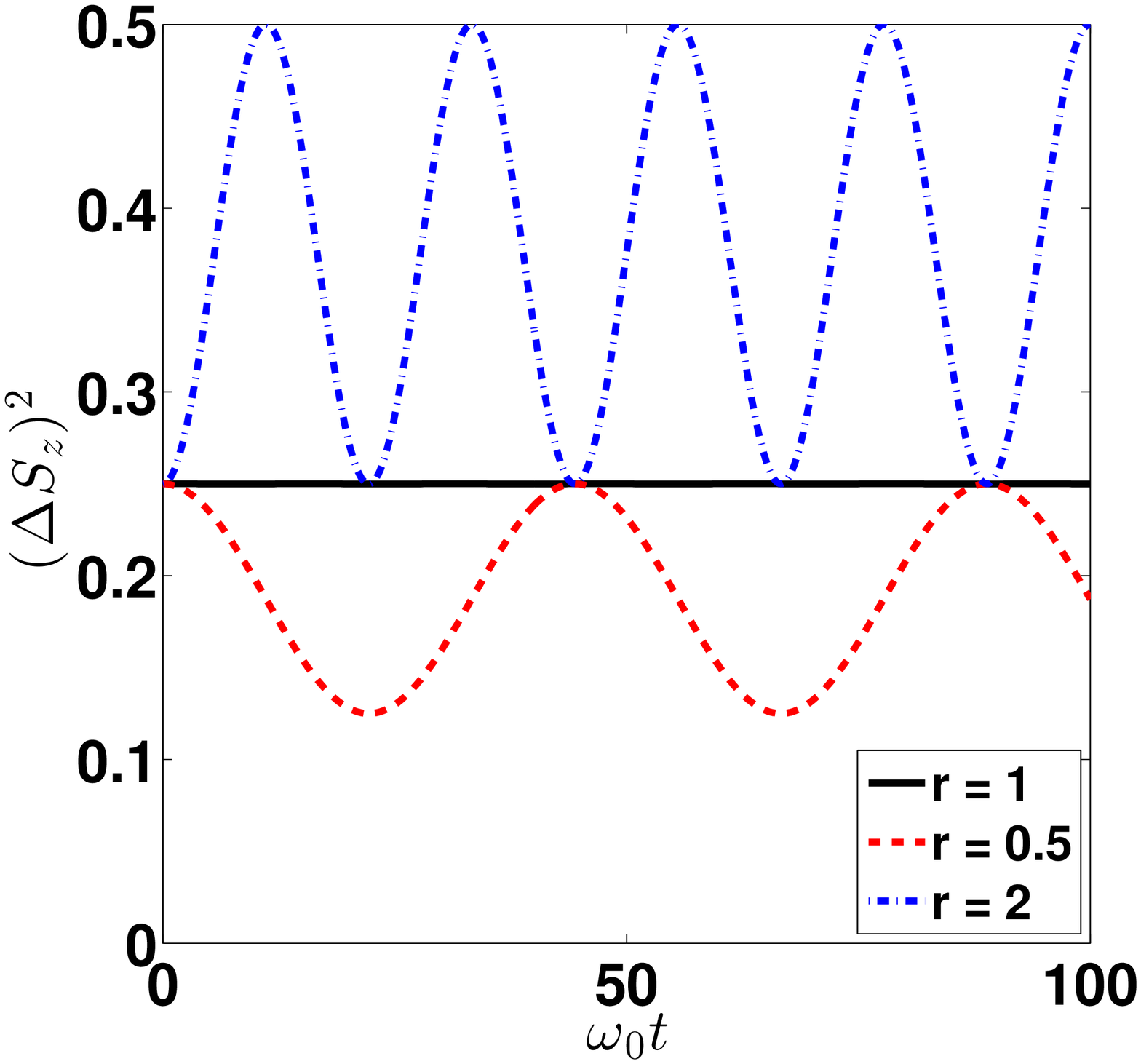}
        }%
	    \subfigure[]{%
            \label{fig:fig2c}
            \includegraphics[width=5.85cm]{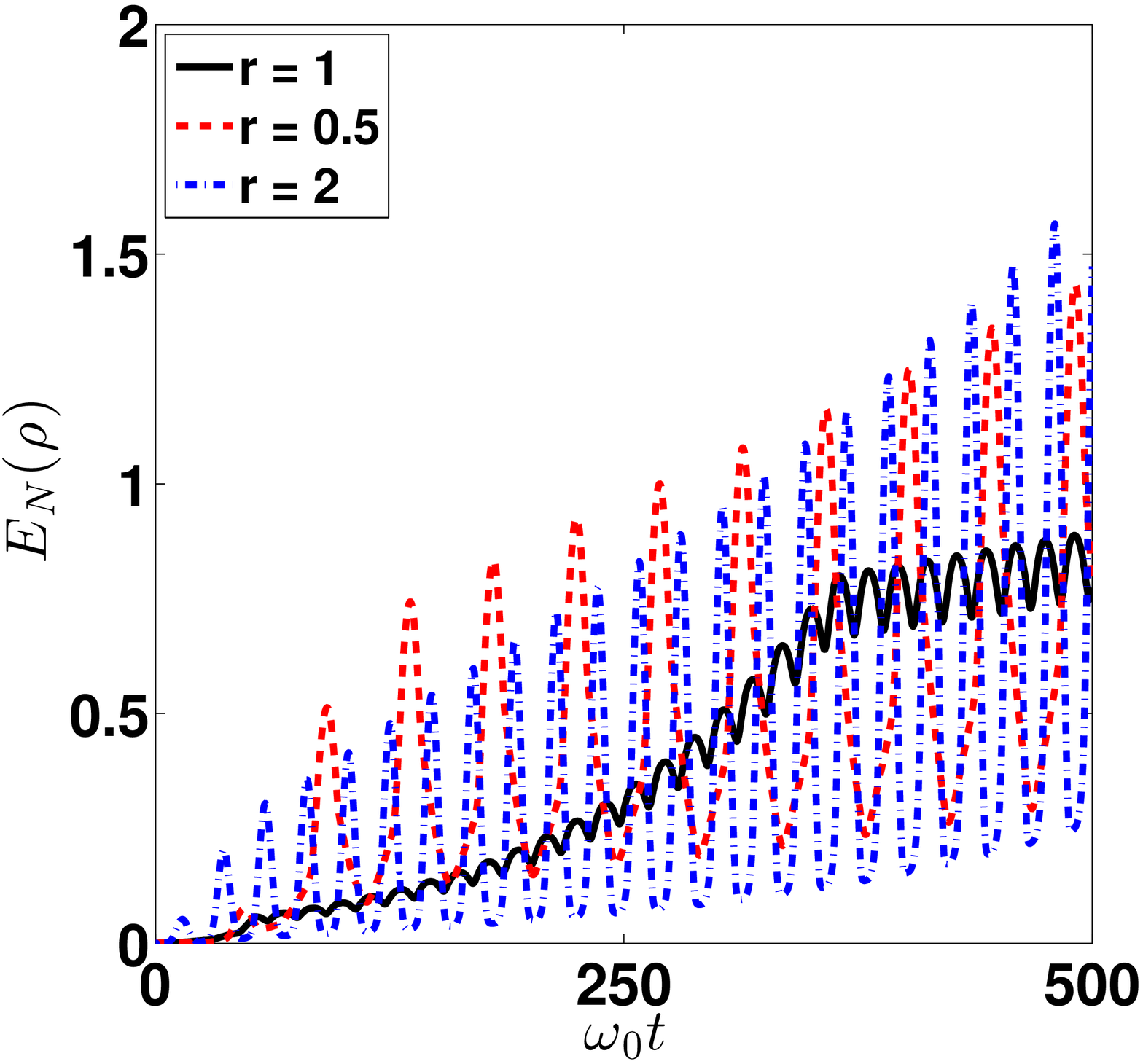}
            }%
    \end{center}
    \caption{%
       Dependence of (a) $\langle S^2\rangle$, (b) $(\Delta S_z)^2$ and (c) $E_N(\rho)$ for $r=1$ (black-solid), $r = 0.5$ (red-dashed) and $r=2$ (blue-dot-dashed) with respect to the scaled time $\omega_0t$ for an initially coherent state $|\psi(0)\rangle=|\alpha\rangle|0\rangle$ with $\alpha=1$.}%
   \label{fig:fig2}
\end{figure*}

We calculate the logarithmic negativity $E_N(\rho)$ to make quantitative discussions on mode entanglement. The logarithmic negativity is a computable and a non-convex entanglement monotone and it is defined as~\cite{plenio2005logarithmic} 
\begin{equation}\label{eq:neg}
E_N(\rho):=log_2||\rho^{T_A}||,
\end{equation}
where $\rho^{T_A}$ stands for the partial transpose with respect to the first subsystem and $||\rho^{T_A}||$ is the trace norm of $\rho^{T_A}$. One important property of the logarithmic negativity is that it does not reduce to the von Neumann entanglement entropy for pure states. It follows that it can detect and measure mode correlations which are not bipartite. Indeed, the absence of bipartite entanglement between cavity modes for Hermitian as well as non-Hermitian cases has been reported~\cite{karakaya}.

There are subtle relations between quantum coherence, correlations, photon localization and delocalization~\cite{hardal2013spin,ferretti2010photon,PhysRevA.83.023805,hardal2014einstein}. If the initial state of the system is a coherent one, then such an interference between non-conservation of the mean number of photons and non-Hermitian dynamics has been also verified~\cite{karakaya}. Here, we shall discuss whether there is an interplay between non-Hermitian dynamics, spin conservation with the associated spin noise and mode correlations as well. 
\begin{figure*}[ht!]
     \begin{center}
            \subfigure[]{%
            \label{fig:fig3a}
            \includegraphics[width=5.85cm]{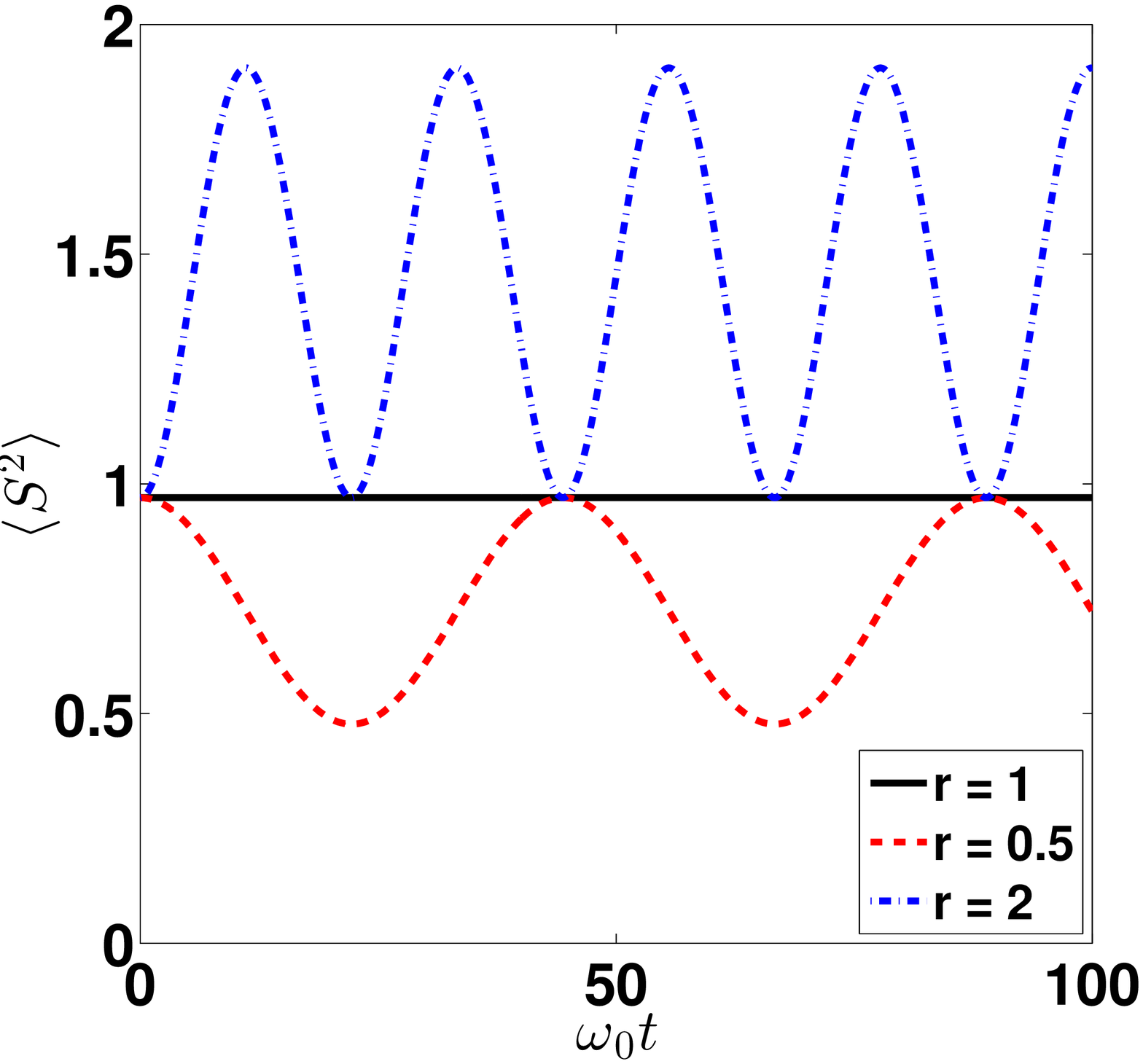}
        }
            \subfigure[]{%
            \label{fig:fig3b}
            \includegraphics[width=5.85cm]{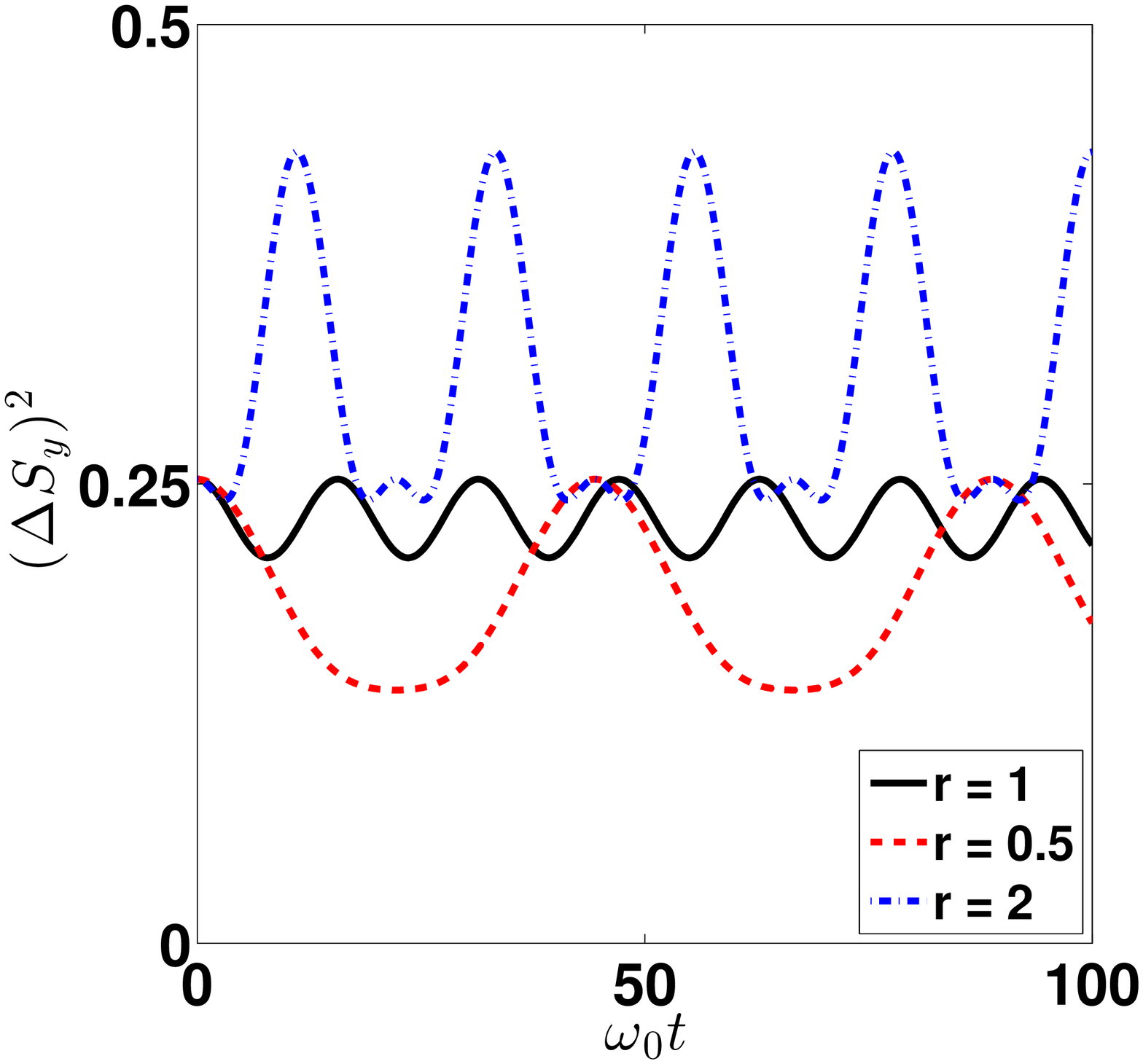}
        }%
	    \subfigure[]{%
            \label{fig:fig3c}
            \includegraphics[width=5.85cm]{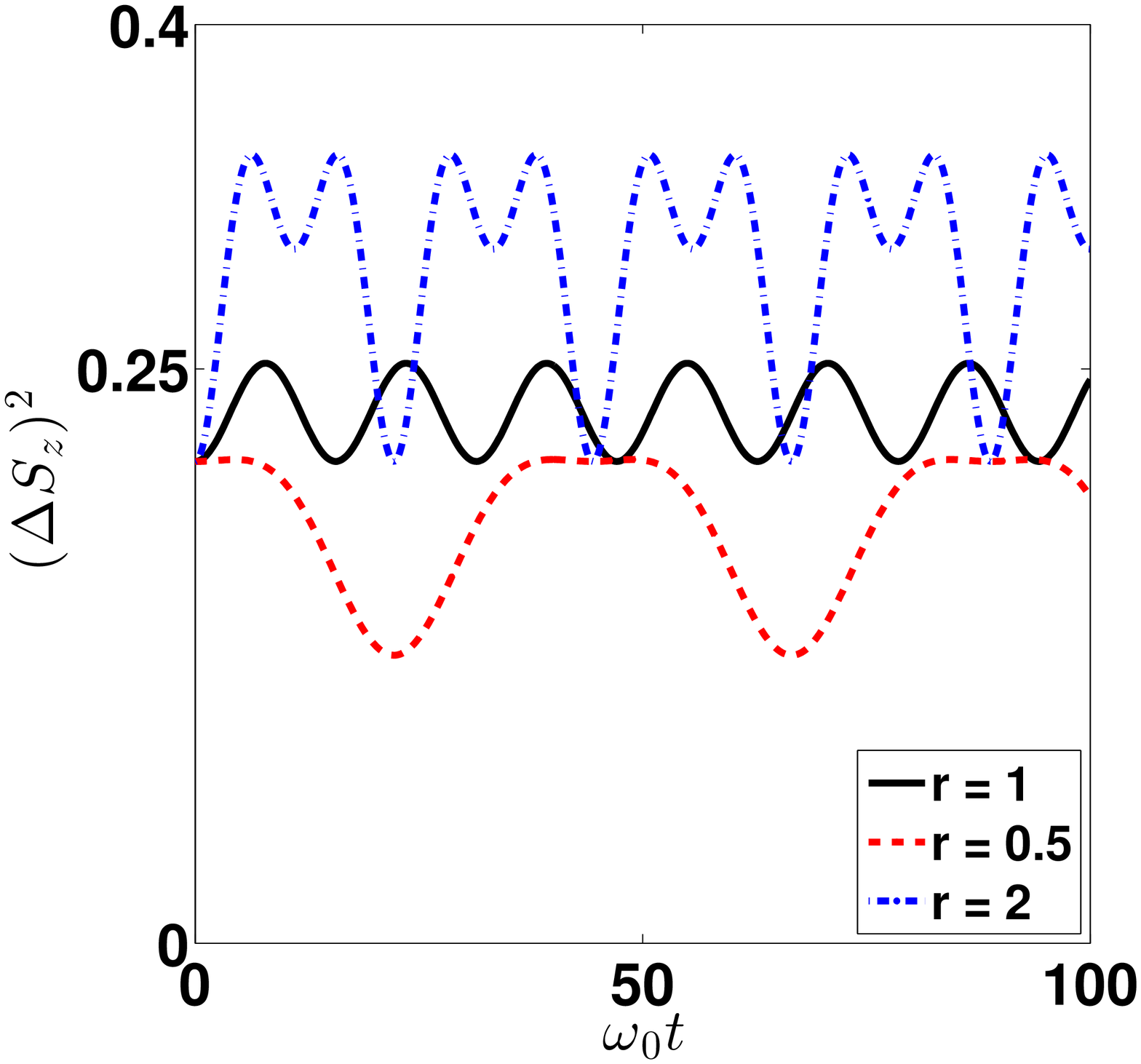}
            }%
    \end{center}
    \caption{%
       Dependence of (a) $\langle S^2\rangle$, (b) $(\Delta S_y)^2$ and (c) $(\Delta S_z)^2$ for $r=1$ (black-solid), $r = 0.5$ (red-dashed) and $r=2$ (blue-dot-dashed) with respect to the scaled time $\omega_0t$ for an initially squeezed state $|\psi(0)\rangle=|\alpha,\epsilon\rangle|0\rangle$ with $\alpha=1$, $\epsilon=0.1$.}%
   \label{fig:fig3}
\end{figure*}
\begin{figure}[h!]
     \begin{center}

            \includegraphics[width=8cm]{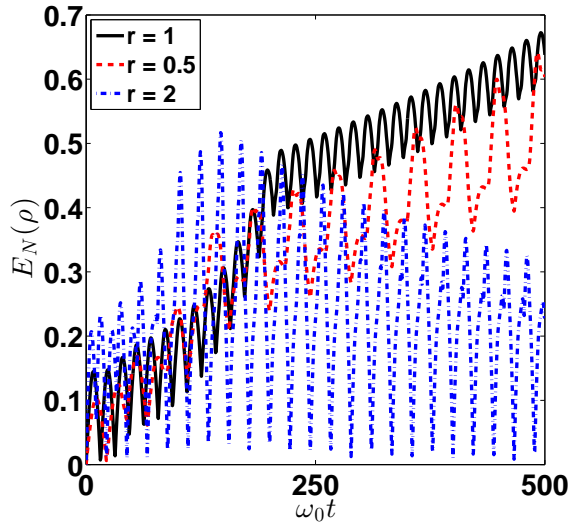}
          
    \end{center}
    \caption{%
       Dependence of $E_N(\rho)$ for $r=1$ (black-solid), $r = 0.5$ (red-dashed) and $r=2$ (blue-dot-dashed) with respect to the scaled time $\omega_0t$ for an initially squeezed state $|\psi(0)\rangle=|\alpha,\epsilon\rangle|0\rangle$ with $\alpha=1$, $\epsilon=0.1$.}%
   \label{fig:fig4}
\end{figure}

\subsection{Initially Fock state}\label{sec:fock} 
We first consider an initial state in which the cavity $A$ is in a Fock state with a single photon whereas the cavity $B$ is in its vacuum
\begin{equation}\label{eq:in_fock}
|\psi(0)\rangle=|1\rangle|0\rangle.
\end{equation}
The mean $\langle N\rangle$ is conserved in both Hermitian and non-Hermitian dynamics~\cite{karakaya}. We numerically verified that the mean of the total spin operator $\langle S^2\rangle=\langle S_x^2+S_y^2+S_z^2\rangle$ is also conserved. Therefore, we can discriminate the effects of non-Hermitian dynamics on the mode correlations with conserved symmetries. 

In Figs.~\ref{fig:fig1a}-\ref{fig:fig1c} we plot the dynamics of the variances $(\Delta S_y)^2$, $(\Delta S_z)^2$ and the logarithmic negativity $E_N(\rho)$ with respect to the scaled time $\omega_0t$, respectively. We calculated that $(\Delta S_x)^2=0.25$ for Hermitian as well as non-Hermitian cases. Fig.~\ref{fig:fig1a} depicts the dynamics of $(\Delta S_y)^2$. When $g_{AB}> g_{BA}$, the photon excitation rate in the empty cavity is faster than the case of $g_{AB}< g_{BA}$, for which the period of the oscillations is bigger than that of the Hermitian case $g_{AB}= g_{BA}$. The variance $(\Delta S_z)^2$ behaves similarly as shown in Fig.~\ref{fig:fig1b}.

Fig.~\ref{fig:fig1c} shows the dynamics of the logarithmic negativity with respect to the scaled time $\omega_0t$. The mode entanglement oscillates between near death $E_N(\rho)\sim0$ and its maximum $E_N(\rho)\sim1$ for all cases. The degree of entanglement mimics the localization-delocalization rate of the photons which depends on the degree of the asymmetry in the coupling strength even if the mean photon number in the system is conserved.  

The model Hamiltonian~(\ref{eq:model1}) can be mapped to that of a two-mode Bose-Einstein condensate (BEC) trapped in a double well~\cite{graefe2008non}. In general, such a system exhibits nonlinear on-site interactions. While the linear tunnelling term between the wells leads to the Josephson oscillations (JO), the nonlinear interactions are responsible for  coherent population self-trapping (ST) of particles. In the present generic model, the absence of the nonlinear interactions leaves the system in the JO regime which results from the single-photon exchange coupling between the cavities. In the JO regime the population imbalance $\langle S_z\rangle$ oscillates around zero with equal amplitudes and never collapses due to the lack of nonlinearity in the system. As expected, these oscillations have residual effects on the dynamics of spin noise as well as the mode entanglement~\cite{choi2005quantum} as reported in Fig.~\ref{fig:fig1}. The large-amplitude oscillations have the period $T$ which is inversely proportional to the non-Hermiticity parameter $r$, i.e., $T\propto1/r$ which shall later be inherited by initially coherent and squeezed state cases as well.
\subsection{Initially coherent state}
Next, we consider an initial state in which the cavity $A$ is in a coherent state with an amplitude of $\alpha=1$ and the cavity $B$ is in its vacuum
\begin{equation}\label{eq:in_coh}
|\psi(0)\rangle=|\alpha\rangle|0\rangle.
\end{equation}
If $g_{AB}\neq g_{BA}$, the mean number of photons $\langle N\rangle$ is not conserved for an initially coherent state~\cite{karakaya}.

In Figs.~\ref{fig:fig2a}-\ref{fig:fig2c}, we plot the dynamics of the mean of the total spin $\langle S^2\rangle$, the variance $(\Delta S_z)^2$ and the logarithmic negativity $E_N(\rho)$ with respect to the scaled time $\omega_0t$, respectively. The variances $(\Delta S_x)^2$ and $(\Delta S_y)^2$ show identical behaviour to that of $(\Delta S_z)^2$ and thus are not presented here. Figure~\ref{fig:fig2a} shows the Hermitian dynamics conserves the mean spin $\langle S^2\rangle$. If $g_{AB}>g_{BA}$, the deviation from the steady value $\langle S^2\rangle=1$ is greater in accordance with the photon number dynamics ~\cite{karakaya}. The associated spin noise behaves similarly and makes negligible oscillations around $(\Delta S_z)^2\sim0.25$ if $g_{AB}=g_{BA}$ as it is expected from a coherent state.

Figure~\ref{fig:fig2c} shows if $g_{AB}=g_{BA}$, the mode entanglement first increase and then starts to oscillate with low amplitudes around $E_N(\rho)\sim0.6$. The coherent trapping of mode entanglement in the JO regime has been also reported for the two-mode BECs~\cite{huang2012fisher}. The non-Hermitian interactions in the cases $g_{AB}>g_{BA}$ and $g_{AB}<g_{BA}$ amplify the mode correlations and have constructive effects in this regard. The number of photons created in the empty cavity differs by the chosen asymmetry in the coupling strengths. On the other hand, the amplification of the mode correlations is mainly a reaction to the broken symmetries as the value of $E_N(\rho)$ is greater than the Hermitian case and almost equal to each other if $g_{AB}<g_{BA}$ or $g_{AB}>g_{BA}$.
\subsection{Initially squeezed state}
We consider an initially squeezed state of the form
\begin{equation}\label{eq:in_sq}
|\psi(0)\rangle=|\alpha, \epsilon\rangle|0\rangle,
\end{equation}
where $\alpha=1$ is the coherent state amplitude and $\epsilon=0.1$ is the squeezing parameter. We numerically verify that the mean $\langle N\rangle$ as well as the number of photons in each cavity shows similar behaviours under Hermitian and non-Hermitian dynamics as reported for initially coherent state ~\cite{karakaya}. The only difference is that for a squeezed state we have
\begin{equation}\label{eq:sq_num}
\langle N\rangle=|\alpha|^2+\sinh^2{\epsilon}.
\end{equation}
In Figs.~\ref{fig:fig3a}-\ref{fig:fig3c}, we plot the dynamics of the mean of the total spin $\langle S^2\rangle$ and the variances $(\Delta S_y)^2$, $(\Delta S_z)^2$ with respect to the scaled time $\omega_0t$, respectively. The variance $(\Delta S_x)^2$ shows identical behaviour as in Fig.~\ref{fig:fig2b} with relatively small amplitudes due to the squeezing and is not presented here. 
Figure~\ref{fig:fig3} shows that the squeezing leads to the reduction of quantum fluctuations in mean spin $\langle S^2\rangle$ as well as in the variances $(\Delta S_y)^2$ and $(\Delta S_z)^2$. 

Figures~\ref{fig:fig3b} and \ref{fig:fig3c} depicts that if $g_{AB}<g_{BA}$, for which the empty cavity is weakly excited~\cite{karakaya}, small plateaus occur where the spin noise is stabilized. If $g_{AB}>g_{BA}$, the empty cavity is strongly exited. In that case, single mode squeezing is not enough to create time intervals in which the noise is rather steady, though it reduces the amplitudes of the fluctuations.

In Fig.~\ref{fig:fig4}, we plot the dynamics of logarithmic negativity with respect to the scaled time $\omega_0t$. If $g_{AB}=g_{BA}$, mode entanglement resembles the dynamics as in the initially coherent state, however it oscillates with relatively higher amplitudes in comply with the spin noise dynamics. If $g_{AB}>g_{BA}$, mode entanglement is blighted by the squeezing. The high amplitude oscillations persist and the maximum value of the logarithmic negativity $E_N(\rho)$ shrinks in comparison with that of the cases of initially coherent and Fock states. Squeezing serves well to the cause in the case of $g_{AB}<g_{BA}$. The amplitude of the oscillations scale down to a pliable level and the coherent entanglement trapping is achieved as in the case of $g_{AB}=g_{BA}$, though the degree of entanglement reduces.
\section{Conclusions}\label{sec:conc}
In summary, we studied the Hermitian and the non-Hermitian dynamics of the mode entanglement in a system of cavities coupled with a chiral mirror. The mode entanglement, characterized by the logarithmic negativity measure, was investigated for initially Fock, coherent and squeezed states. 

For an initially Fock state both the total number of photons~\cite{karakaya} and the mean of the total spin are conserved regardless of the type of the dynamics. The single photon exchange is a delocalizing and mode correlating interaction~\cite{hardal2013spin}. As a result, the period of oscillations in the time evolution of the mode entanglement mimics that of the photon exchange but keeps the degree of entanglement constant. The former is also inherited by the initially coherent and squeezed state cases as well.

The interplay between coherence, correlations and the non-conservation of mean spin as well as mean number of photons is revealed in the case of an initially coherent state. The degree of mode entanglement is amplified if the coupling between the two cavity is non-reciprocal. The amplification is nearly equal whether $g_{AB}>g_{BA}$ or $g_{AB}<g_{BA}$ whereas the number of photons are quite different depending on the asymmetries. 

Lastly, we considered an initially squeezed state to diminish the amplitudes of the oscillations in the dynamics of the mode entanglement. We found that if the empty cavity is weakly excited squeezing leads to the desired reduction  with the expense in the magnitude of the entanglement.

Our results demonstrate that the non-reciprocal exchange interactions may be used to ensure an effective control over the dynamics as well as the degree of the quantum entanglement which could be desirable from the perspective of quantum information technologies.

\acknowledgements
We are grateful to  \"{O}zg\"{u}r  E. M\"ustecapl{\i}o\u{g}lu for his encouragement and illuminating discussions. We thank Hakan E. T\"{u}reci for fruitful discussions and
Princeton University for their hospitality. 
\bibliographystyle{apsrev}
\bibliography{non_her}
\end{document}